\documentclass[pre,twocolumn,preprintnumbers,amsmath,amssymb,nofootinbib,superscriptaddress]{revtex4}

\usepackage{graphicx,bm,float}

\makeatletter
\def\graphicscale{\twocolumn@sw{0.3}{0.4}}
\def\graphicthreescale{\twocolumn@sw{0.3}{0.4}}

\begin{document}

\title{Higher-charge three-dimensional compact lattice Abelian-Higgs models}

\author{Claudio Bonati} 
\affiliation{Dipartimento di Fisica dell'Universit\`a di Pisa
        and INFN, Largo Pontecorvo 3, I-56127 Pisa, Italy}

\author{Andrea Pelissetto}
\affiliation{Dipartimento di Fisica dell'Universit\`a di Roma Sapienza
        and INFN, Sezione di Roma I, P.le A. Moro 2, I-00185 Roma, Italy}

\author{Ettore Vicari} 
\affiliation{Dipartimento di Fisica dell'Universit\`a di Pisa
        and INFN, Largo Pontecorvo 3, I-56127 Pisa, Italy}

\date{\today}

\begin{abstract}
We consider three-dimensional higher-charge multicomponent lattice
Abelian-Higgs (AH) models, in which a compact U(1) gauge field is
coupled to an $N$-component complex scalar field with integer charge
$q$, so that they have local U(1) and global SU($N$) symmetries.  We
discuss the dependence of the phase diagram, and the nature of the
phase transitions, on the charge $q$ of the scalar field and the
number $N\ge 2$ of components.  We argue that the phase diagram of
higher-charge models presents three different phases, related to the
condensation of gauge-invariant bilinear scalar fields breaking the
global SU($N$) symmetry, and to the confinement/deconfinement of
external charge-one particles.  The transition lines separating the
different phases show different features, which also depend on the
number $N$ of components.  Therefore, the phase diagram of
higher-charge models substantially differs from that of unit-charge
models, which undergo only transitions driven by the breaking of the
global SU($N$) symmetry, while the gauge correlations do not play any
relevant role.  We support the conjectured scenario with numerical
results, based on finite-size scaling analyses of Monte Carlo
simuations for doubly-charged unit-length scalar fields with small and
large number of components, i.e. $N=2$ and $N=25$.
\end{abstract}

\maketitle


\section{Introduction}
\label{intro}

Abelian U(1) gauge theories with multicomponent scalar fields,
characterized by a global SU($N$) symmetry, emerge as effective
theories in many different physical contexts ~\cite{HLM-74,%
  DH-81,FH-96,MHS-02,NRR-03,RS-90,TIM-05,TIM-06,Kaul-12,KS-12,%
  BMK-13,NCSOS-15,WNMXS-17}. In particular, they provide an effective
description of deconfined quantum critical points~\cite{SBSVF-04}, for
example, of the N\'eel to valence-bond-solid transition in
two-dimensional antiferromagnetic SU(2) quantum
systems~\cite{Sandvik-07,MK-08,JNCW-08,Sandvik-10,%
  HSOMLWTK-13,CHDKPS-13,PDA-13,SGS-16}. These quantum models and their
classical counterparts have been extensively studied to understand
their different phases and the nature of their phase transitions.  A
crucial role is played by topological aspects, like the Berry phase or
the compact/noncompact nature of the gauge fields.  For example, the
critical behavior of the lattice CP$^{N-1}$ model, which is the
simplest classical model with U(1) gauge symmetry, depends on the
presence/absence of topological
defects~\cite{MS-90,MV-04,NCSOS-11,NCSOS-13,PV-20-mfcp}, such as
monopoles, both for $N=2$ and large values of $N$.  Analogous
differences emerge in the behavior of compact and noncompact lattice
formulations of scalar electrodynamics, i.e., of the multicomponent
Abelian-Higgs model, see, e.g., Refs.~\cite{SBSVF-04, KPST-06,
  Sandvik-07, MK-08, JNCW-08, MV-08, KMPST-08-a, KMPST-08, CAP-08,
  LSK-09, CGTAB-09, CA-10, BDA-10, Sandvik-10, HBBS-13, Bartosch-13,
  HSOMLWTK-13, CHDKPS-13, PDA-13, BS-13, NCSOS-15, NSCOS-15, SP-15,
  SGS-16, PV-20-mfcp, SN-19, SZ-20,PV-19-cah,BPV-20-ncah}.

In this paper we consider three-dimensional (3D) higher-charge
multicomponent lattice Abelian-Higgs (AH) models. In these models a
compact U(1) gauge field is coupled to an $N$-component complex scalar
field with integer charge $q > 1$, so that they are invariant under
local U(1) and global SU($N$) transformations.  We study the
dependence of the thermodynamic properties, such as the phase diagram
and the nature of the phase transitions, on the value of the charge
$q$.  Our work extends previous studies of the compact lattice AH
model with a single ($N=1$) higher-charge complex scalar
field~\cite{FS-79,SSSNH-02,SSNHS-03,NSSS-04,CFIS-05,CIS-06,WBJS-08} to
multicomponent $N\ge 2$ theories. For $N=1$ and $q\ge 2$, the phase
diagram is characterized by two phases, that are distinguished by the
confinement/deconfinement of single-charge external particles. In this
case, the Wilson loops associated with charge-one particles can be
considered as the order parameter of the transitions, which separate
the confined high-temperature phase, in which Wilson loops obey the
area law, from the deconfined phase.  As we shall see, for $N\ge 2$
the confinement/deconfinement of charge-one external sources also
plays an important role in determining the phase diagram. However,
there are also new features related to the breaking of the global
SU($N$) symmetry.

The phase diagram of the multicomponent compact AH model with $q \ge
2$ also presents notable differences with respect to that of the same
model with $q=1$.  For $N\ge 2$ and $q=1$ (this has been explicitly
verified in the London limit in which the scalar fields have unit
length) there are only two phases, that can be characterized by using
a gauge-invariant scalar-field order parameter, while gauge
fluctuations do not play any relevant role~\cite{PV-19-cah}.  In
particular, for $N=2$ the transition line between the high- and the
low-temperature phase shows continuous transitions belonging to the
CP$^{1}$---equivalently, O(3) vector---universality class.

\begin{figure}[tbp]
\includegraphics*[width=1.0\columnwidth]{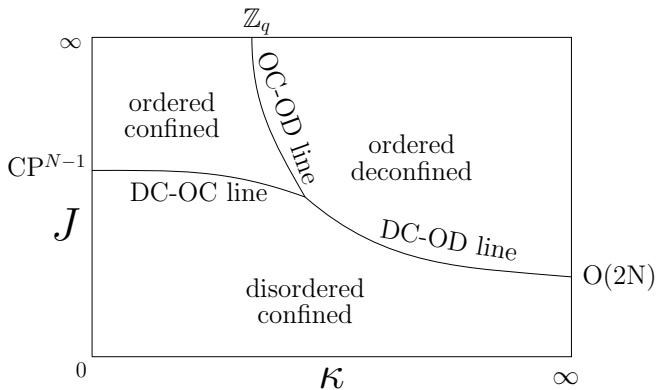}
  \caption{Sketch of the $J$-$\kappa$ phase diagram of the 3D
    multicomponent lattice Abelian-Higgs model, in which a compact
    U(1) gauge field is coupled to an $N$-component unit-length
    complex scalar field with charge $q\ge 2$, for generic $N\ge
    2$. The Hamiltonian parameter $J$ is associated with the kinetic
    gauge-invariant term of the scalar field, while $\kappa$
    represents the inverse gauge coupling.  See text for a description
    of the various phases and transition lines.  We also report the
    models emerging in some limiting cases: the CP$^{N-1}$ model for
    $\kappa=0$, the O($2N$) vector model for $\kappa\to\infty$, and
    the lattice ${\mathbb Z}_q$ gauge model for $J\to\infty$.  }
\label{phdiasketch}
\end{figure}

We argue, and present numerical results to support our arguments,
that, for $q\ge 2$ and $N\ge 2$, the model with compact gauge fields
and unit-length complex scalar fields (London limit) shows three
phases, as sketched in Fig.~\ref{phdiasketch}.  They are characterized
by the behavior of the gauge-invariant correlations of the scalar
fields, which may give rise to the breaking of the SU($N$) global
symmetry, and by the confinement/deconfinement of charge-one external
particles, that is signalled by the large-size behavior of the Wilson
loops of the gauge fields.  As shown in Fig.~\ref{phdiasketch}, for
small $J$ and any $\kappa \ge 0$, there is a phase in which
scalar-field correlations are disordered and single-charge particles
are confined (the Wilson loop obeys the area law).  For large values
of $J$ (low-temperature region) scalar correlations are ordered and
the SU($N$) symmetry is broken. Two phases occur here: for small
$\kappa$, single-charge particles are confined, while they are
deconfined for large $\kappa$.  These phases are separated by three
transition lines meeting at a multicritical point: the DC-OD
transition line between the disordered/confined (DC) and the
ordered/deconfined (OD) phases, the DC-OC line between the
disordered/confined and ordered/confined (OC) phases, and the OC-OD
line between the ordered/confined and ordered/deconfined phases.

The three transition lines have different features, since they are
associated with different phases.  Moreover, their nature crucially
depends on the number $N$ of components.  In particular, for $q=2$ and
$N=2$, we provide evidence that the transitions along the DC-OC and
OC-OD lines are continuous, belonging to the O(3) vector and Ising
universality class, respectively.  The transitions along the DC-OD
line are of first order.  For large values of $N$ we expect a
different behavior along the DC-OC and DC-OD lines.  The transitions
along the DC-OC line are expected to be first order.  As we shall see,
our numerical results for $q=2$ and $N=25$ provide evidence of
continuous transitions along the DC-OD line.

We note that the different qualitative behavior of the models with
$q=1$ and $q\ge 2$ is a specific feature of the compact formulation of
the AH theory.  Indeed, in the AH model with noncompact gauge fields a
change of the charge of the scalar field is equivalent to a change of
the strength of the gauge coupling. Therefore, apart from a trivial
rescaling of the gauge coupling, the phase diagram is the same. We
also note that the noncompact formulation of the AH theory should be
recovered in the $q\to\infty$ limit of the compact formulation, with
an appropriate correspondence of the gauge couplings.

The paper is organized as follows.  In Sec.~\ref{sec2} we introduce
the lattice AH model with an $N$-component scalar field of generic
charge $q$, and define the relevant observables that characterize the
phase transitions.  In Sec.~\ref{phadia} we present the possible
scenarios for the phase diagram and for the nature of the transition
lines. Sec.~\ref{numres} presents our numerical results for $q=2$: we
report FSS analyses~\cite{FB-72,Barber-83,Privman-90,PV-02} of the
Monte Carlo (MC) results for $N=2$ and $N=25$, which allow us to
determine the phase diagram of the model and to characterize the
different transition lines.  In Sec.~\ref{monsup} we discuss the role
that monopoles play in the compact model with $q\ge 2$.  Finally, we
draw our conclusions in Sec.~\ref{conclu}.

\section{The  higher-charge lattice AH model} \label{sec2}

\subsection{The model}
\label{model}

We consider a three-dimensional lattice AH model, in which the scalar
field ${\bm z}_{\bm x}$ is a complex $N$-component unit vector
($\bar{\bm z}_{\bm x}\cdot{\bm z}_{\bm x} = 1$) of integer charge $q$
defined on the sites ${\bm x}$ of a cubic lattice.  For the gauge
fields, we use the compact Wilson formulation, associating complex
variables $\lambda_{{\bm x},\mu}$ with $|\lambda_{{\bm x},\mu}| = 1$
to each link connecting the site ${\bm x}$ with the site ${\bm
  x}+\hat\mu$, where $\hat\mu=\hat{1},\hat{2},\hat{3}$ are unit
vectors along the lattice directions.  The Hamiltonian reads
\begin{equation}
H = J N H_z + \kappa H_g \,.
\label{qAHmodel}
\end{equation}
The first term is the interaction term for the scalar fields of charge
$q$:
\begin{eqnarray}
H_z = - \sum_{{\bm x}, \mu} \left( \bar{\bm{z}}_{\bm x} \cdot
\lambda_{{\bm x},\mu}^q\, {\bm z}_{{\bm x}+\hat\mu} + {\rm
  c.c.}\right)\,,
\label{HamZ}
\end{eqnarray}
where the sum is over all lattice links of the cubic lattice.  The
second term is the usual Wilson Hamiltonian for a U(1) gauge field:
\begin{eqnarray}
H_g = 
- \sum_{{\bm x},\mu<\nu} \left(\Pi_{{\bm x},\mu\nu} + {\rm c.c.}\right) \,,
\label{Hamg} 
\end{eqnarray}
where the sum is over the lattice plaquettes and $\Pi_{{\bm
    x},\mu\nu}$ is the field strength associated with each plaquette,
\begin{equation}
\Pi_{{\bm x},\mu\nu} = 
\lambda_{{\bm x},{\mu}} \,\lambda_{{\bm x}+\hat{\mu},{\nu}}
\,\bar{\lambda}_{{\bm x}+\hat{\nu},{\mu}}
  \,\bar{\lambda}_{{\bm x},{\nu}} \, .
\label{plaquette}
\end{equation}
For any integer $q$ the model is invariant under the global SU($N$)
transformations $z_{\bm x} \to U z_{\bm x}$, $\lambda_{{\bm x},\mu}
\to \lambda_{{\bm x},\mu}$, where $U\in {\rm SU}(N)$, and under the
local U(1) transformations $z_{{\bm x}} \to e^{iq\phi_{\bm x} } z_{\bm
  x}$, $\lambda_{{\bm x},\mu} \to e^{i(\phi_{\bm x} -\phi_{{\bm x} +
    \hat{\mu}}) } \lambda_{{\bm x},\mu}$, where $\phi_{\bm x}$ is a
site-dependent phase. For $q>1$ there are additional global symmetries
that only involve the gauge field.  If we consider the sites ${\bm y}$
that belong to a given plane---for definiteness consider a plane
orthogonal to the direction $\hat{1}$--- the Hamiltonian is invariant
under the transformation $\lambda_{{\bm y},1} \to \alpha \lambda_{{\bm
    y},1}$ on all these sites, where $\alpha$ satisfies
$\alpha^q=1$. The $\mathbb{Z}_q$ symmetry is the analogue of the
center symmetry that is present in pure lattice gauge theories.  Its
spontaneous breaking signals the deconfinement of the single-charge
sources. The partition function of the system reads
\begin{equation}
Z = \sum_{\{{\bm z}\},\{\lambda\}} e^{-\beta H}\,.
\label{partfun}
\end{equation}
In the following we rescale $J$ and $\kappa$ by $\beta$, thus formally
setting $\beta=1$.

\subsection{Observables}
\label{observables}

To characterize phase transitions associated with the breaking of the
SU($N$) symmetry, we consider correlations of the gauge-invariant
hermitean operator
\begin{equation}
Q^{ab}_{\bm x} = \bar{z}_{\bm x}^a z_{\bm x}^b - {1\over N}
\delta^{ab}
\label{qdef}.
\end{equation}
Its two-point correlation function is defined as
\begin{equation}
G({\bm x}-{\bm y}) = \langle {\rm Tr}\, Q_{\bm x} Q_{\bm y} \rangle\,,  
\label{gxyp}
\end{equation}
where the translation invariance of the system has been taken into
account.  The susceptibility and the correlation length are defined as
$\chi=\sum_{\bm x} G({\bm x})$ and
\begin{eqnarray}
\xi^2 \equiv  {1\over 4 \sin^2 (\pi/L)}
{\widetilde{G}({\bm 0}) - \widetilde{G}({\bm p}_m)\over 
\widetilde{G}({\bm p}_m)}\,,
\label{xidefpb}
\end{eqnarray}
where $\widetilde{G}({\bm p})=\sum_{{\bm x}} e^{i{\bm p}\cdot {\bm x}}
G({\bm x})$ is the Fourier transform of $G({\bm x})$, and ${\bm p}_m =
(2\pi/L,0,0)$ is the minimum nonzero lattice momentum.

In our analysis we will consider renormalization-group (RG)
invariant quantities,
such as $R_\xi = \xi/L$ and the Binder parameter
\begin{equation}
U = {\langle \mu_2^2\rangle \over \langle \mu_2 \rangle^2} \,, \qquad
\mu_2 = 
\sum_{{\bm x},{\bm y}} {\rm Tr}\,Q_{\bm x} Q_{\bm y}\,.
\label{binderdef}
\end{equation}
We also consider the energy-related observables
\begin{eqnarray}
&& 
E_z = -{1\over 2V} \langle H_z\rangle\,,\quad
C_z ={1\over 4V} \left( \langle H_z^2 \rangle - \langle H_z  \rangle^2\right),
\label{eczdef} \\
&& 
E_g = -{1\over 2V} \langle H_g\rangle\,,\quad
C_g ={1\over 4V} \left( \langle H_g^2 \rangle - \langle H_g  \rangle^2\right),
\qquad
\label{ecgdef}
\end{eqnarray}
where $V=L^3$. They correspond to the energy density and to the
specific heat associated with the scalar and gauge part of the
Hamiltonian, respectively.  The normalizations have been chosen so
that $E_z$ and $E_g$ converge to 3 in the ordered limit
$J,\,\kappa\to\infty$.  We also consider the third cumulant
\begin{equation}
H_{g3} = \langle (H_g - \langle H_g\rangle)^3 \rangle\,.
\end{equation}
We consider correlation functions of plaquette operators,
such as the connected correlation function 
\begin{eqnarray}
G_{\pi}({\bm x} - {\bm y}) = \langle \Sigma_{\bm x} \Sigma_{\bm y}\rangle_c\,,
\quad
\Sigma_{\bm x} = \hbox{Re} \, \sum_{\mu<\nu} \Pi_{{\bm x},\mu\nu}\,.
\label{plaope}
\end{eqnarray}
Note that $\sum_{\bm x} G_{\pi}({\bm x}) = C_g$,
cf. Eq.~(\ref{ecgdef}).  Using $G_{\pi}({\bm x})$ one can also define
a correlation length $\xi_{\pi}$ as in Eq.~(\ref{xidefpb}).

We finally define a field-strength correlation function.  For
definiteness, we select the $x$ direction and define ($x,y,z$ are the
coordinates of each lattice point)
\begin{equation}
 n_x = \hbox{Im}\, \left(\sum_{yz} \Pi_{(x,y,z),23} \right).
\end{equation}
Then we consider a lattice plane orthogonal to the unit vector
$\hat{1}$ and sum over all plaquettes that belong to the plane. When
the gauge fields are close to 1, $n_x$ is essentially the sum of the
field strengths on the plane. Then, we define a correlation function
\begin{equation}
G_F(x_1 - x_2) = \langle n_{x_1} n_{x_2} \rangle,
\end{equation}
and a correlation length $\xi_F$ using Eq.~(\ref{xidefpb}). Of course,
one can define an analogous correlation function $G_F(x)$ by
considering the $y$ or the $z$ direction. In our simulation, we
compute $G_F(x)$ by averaging the correlation function over the three
lattice directions.

\section{The phase diagram}
\label{phadia}

\subsection{Some limiting cases}
\label{limcases}

To understand the phase diagram of the model, it is useful to consider
some particular cases, in which the thermodynamic behavior is already
known. No transitions are expected along the $J=0$ line, while
transitions occur along the $\kappa=0$, the $J=\infty$, and the
$\kappa=\infty$ lines.

\subsubsection{Phase diagram along the $\kappa=0$ line.}  

For $\kappa=0$ the model is equivalent to a lattice formulation of the
CP$^{N-1}$ models with explicit lattice gauge
variables~\cite{PV-19-cpn}.  Indeed, for $\kappa=0$ the charge $q$
does not play any role: one can redefine $\lambda^q \to \lambda'$,
without changing the model.  Its phase diagram has two phases
separated by a finite-temperature transition, where the order
parameter is the gauge-invariant bilinear operator defined in
Eq.~(\ref{qdef}). Some estimates of the transition point $J_c$ have
been obtained in Refs.~\cite{PV-19-cpn,PV-20-largeN}, and summarized
in Ref.~\cite{BPV-20-ncah}.  For $N=2$ the transition is continuous,
belonging to the O(3) vector universality class (accurate estimates of
the O(3) critical exponents can be found in
Refs.~\cite{Hasenbusch-20,KP-17,HV-11,PV-02,CHPRV-02,GZ-98}). For
$N\ge 3$ it is instead of first order: it is weak for
$N=3$~\cite{PV-19-cpn} and becomes stronger and stronger with
increasing $N$~\cite{PV-20-largeN}. We expect that nature of the
transitions to persist for finite, sufficiently small values of
$\kappa$.

\subsubsection{Phase diagram along the $J=\infty$ line. }

For $J\to\infty$ the relevant configurations are those that minimize
the spin Hamiltonian $H_z$ defined in Eq.~(\ref{HamZ}). In the
lowest-energy configuration the fields $z_{\bm x}$ satisfy
\begin{equation}
z_{\bm x} = \lambda_{{\bm x},\mu}^q z_{{\bm x} + \hat{\mu}}.
\end{equation}
Iterating this relation along a plaquette, we obtain 
\begin{equation}
  \Pi_{{\bm x},\mu\nu}^q = 1
\end{equation}
on all plaquettes. A similar result holds for the product of the
fields $\lambda_{{\bm x},\mu}^q$ along any topologically nontrivial
lattice loop.  Therefore, modulo gauge transformations, we can set
$\lambda_{{\bm x},\mu}^q = 1$ on any lattice link. This condition
implies
\begin{equation}
\lambda_{{\bm x},\mu} = \exp\left({2 \pi i n/q}\right), \qquad
   n = 0,\ldots ,q-1.
\end{equation}
Thus, in the $J\to\infty$ limit we obtain a lattice ${\mathbb Z}_q$
gauge theory, in which the spin variables are associated with the
links of the lattice.  These models can be related by duality to
$q$-state clock spin models~\cite{SSNHS-03}, characterized by a global
${\mathbb Z}_q$ symmetry. For $q=2$, the $q$-state clock model is
equivalent to the standard Ising model and thus we expect an Ising
transition for $J=\infty$. For $q=3$, the $q$-state clock model is
equivalent to a three-state Potts model, which can only undergo
first-order transitions.  For larger values of $q$, we expect a
continuous transition.  It belongs to the Ising universality class for
$q=4$~\cite{HS-03}, and to the 3D $XY$ universality class for $q\ge
5$~\cite{HS-03,Hasenbusch-19,PSS-20}. It is important to note that,
for $q=2$, duality allows us to map the ${\mathbb Z}_2$ gauge
Hamiltonian (\ref{Hamg}) onto the usual nearest-neighbor Ising
model. This allows us to predict $\kappa_c = {1\over 4} \ln \coth
\beta_{I,c}$, where $\beta_{I,c}$ is the inverse temperature of the
Ising model.  Using ~\cite{FXL-18} $\beta_{I,c} = 0.221654626(5)$, we
obtain $\kappa_c = 0.380706646(6)$.

\subsubsection{Phase diagram along the $\kappa=\infty$ line.}  

For $\kappa\to \infty$ the gauge degrees of freedom are frozen and we
can set $\lambda_{{\bm x},\mu} = 1$ modulo gauge transformations.  In
this limit the model is therefore equivalent to the O($2N$) vector
model, which undergoes a continuous transitions for any $N$. The same
occurs in the standard lattice $q=1$ AH models with compact and
noncompact gauge fields, see, e.g.,
Refs.~\cite{PV-19-cah,BPV-20-ncah}.  Estimates of the critical values
$J_c$ along the $\kappa=\infty$ line are summarized in
Ref.~\cite{BPV-20-ncah}. They are obtained from the results reported
in Refs.~\cite{BFMM-96,BC-97,DPV-15,PV-20-largeN,CPRV-96}.  The RG
analysis of the continuum AH field theory, see, e.g.,
Refs.~\cite{PV-19-cah,BPV-20-ncah}, predicts that gauge modes are a
relevant perturbation of the O($2N$) fixed point.  Therefore, the
DC-OD transitions do not belong to the O($2N$) vector universality
class.  However, the O($2N$) continuous transition for $\kappa=\infty$
gives rise to crossover phenomena for large values of $\kappa$.

\subsection{Transition lines}
\label{trlines}

On the basis of the above considerations, the most natural phase
diagram is the one reported in Fig.~\ref{phdiasketch}. There are three
different phases, characterized by the behavior of the spin and gauge
correlations. The spin order parameter is the bilinear scalar operator
$Q_{\bm x}$ defined in Eq.~(\ref{qdef}), that signals the breaking of
the global SU($N$) symmetry. The behavior of the gauge modes can be
understood by looking at the behavior of the charge-one Wilson loops
\begin{equation}
W_{\cal C} = \prod_{\ell\in \cal C} \lambda_\ell,
\end{equation}
where $\cal C$ is a closed lattice loop. Depending on the phase, the
Wilson loop may or may not obey the area law; correspondingly
charge-one sources may be confined or deconfined.

Using these two order parameters, we can characterize the different
phases.  For small $J$ and any $\kappa$, there is a high-temperature
disordered-confined (DC) phase with disordered scalar-field
correlations and confined charge-one particles.  For large $J$ there
are two phases, in which spin correlations are ordered and the global
SU($N$) symmetry is broken. They differ in the behavior of the gauge
correlations: for small $\kappa$ charge-one sources are confined,
while for large $\kappa$ they are deconfined.  These phases are
separated by three distinct transition lines, that presumably meet at
a multicritical point (MCP) at ($\kappa_{m}, J_{m})$, see
Fig.~\ref{phdiasketch}.

Note that the value of the charge plays here a crucial role. For the
unit-charge theory, the area law never holds, as soon as the scalar
interaction is turned on, a phenomenon known as screening. For
instance, for $J$ small there is always a contribution to $W_{\cal C}$
of order $J^{p_{\cal C}}$, where $p_{\cal C}$ is the length of the
Wilson loop. These contributions are absent for any $q\ge 2$.

The above predictions strictly apply to the lattice AH model in the
London limit, in which the modulus $|z_{\bm x}|$ is kept
fixed. However, we believe that the same phases and transitions occur
in more general models in which the constraint is relaxed.

\subsubsection{The OC-OD transition line}

The nature of the OC-OD transition line is expected to be independent
of the number of components, at least for sufficiently large values of
$J$.  For $q=2$ it is continuous, belonging to the Ising universality
class, for any $J > J^*$, where $J^*$ may coincide with the position
of the MCP, i.e., $J^*\ge J_{m}$.  Along this transition line the
scalar-field fluctuations are expected to be irrelevant, acting as
spectators.  For $q=3$ the first-order nature of the transition at
$J=\infty$ is expected to persist for finite values of $J$.  The
transition line is expected to be continuous for $q\ge 4$ and to
belong to the $XY$ universality class for $q\ge 5$.

\subsubsection{The DC-OC transition line}

The transitions along the DC-OC line are expected to have the same
nature as the $\kappa=0$ transition, at least for $\kappa < \kappa^*$,
where $\kappa^*$ must satisfy $\kappa^* \le \kappa_{m}$.  As we said,
we do not expect the addition of $H_g$ for small $\kappa$ to change
the nature of the transition. Therefore, as it occurs for $\kappa =
0$, we expect gauge modes to be irrelevant.  Thus, the transitions for
$N=2$ should be continuous and belong to the O(3) vector universality
class, while they should be of first order for $N\ge 3$. As in the
CP$^{N-1}$ model ($\kappa=0$), the DC-OC transition line is
characterized by the condensation of the gauge-invariant bilinear
operator (\ref{qdef}).

\subsubsection{The DC-OD transition line}

The nature of the DC-OD transition line is less clear.  The bilinear
operator (\ref{qdef}) is again expected to be an appropriate order
parameter. Moreover, as $J$ increases across the transition line, we
also expect deconfinement: the area law ceases to hold in the OD
phase.  On the basis of the numerical results reported in
Sec.~\ref{numres}, we shall argue that the behavior along the DC-OD
transition line is controlled by the stable fixed point of the
continuum AH field theory, whose Lagrangian reads
\begin{equation}
{\cal L} = 
|D_\mu{\bm\Phi}|^2
+ r\, {\bm \Phi}^*{\bm \Phi} + 
\frac{1}{6} u \,({\bm \Phi}^*{\bm \Phi})^2 + 
\frac{1}{4 g^2} \,F_{\mu\nu}^2 
\,,
\label{abhim}
\end{equation}
where $\Phi$ is a $N$-component complex scalar field,
$F_{\mu\nu}\equiv \partial_\mu A_\nu - \partial_\nu A_\mu$, and $D_\mu
\equiv \partial_\mu + i A_\mu$.  This is analogous to what occurs
along the Coulomb-Higgs line of the lattice AH model with unit-charge
spin fields and noncompact gauge fields~\cite{BPV-20-ncah}.  We recall
that the RG flow of the AH field theory predicts that continuous
transitions may be observed only for $N >
N_c$~\cite{HLM-74,MZ-03,IZMHS-19}, with $4 < N_c < 10$ in three
dimensions~\cite{IZMHS-19,BPV-20-ncah}.

\subsection{Nature of the bicritical point for $q=2$ and
$N=2$}
\label{multi}

As discussed above, the phase diagram of the lattice AH model with
charge $q= 2$ is characterized by three transition lines meeting at a
MCP, which is usually called bicritical
point~\cite{LF-72,FN-74,NKF-74,CPV-03}. To discuss its nature, it is
crucial to identify the relevant critical modes.  Let us focus on the
particular case $N=2$, where we expect continuous transitions along
the DC-OC line with an O(3) scalar order parameter, and continuous
transitions along the OC-OD line with an Ising order parameter.
Therefore, the nature of the MCP is determined by the competition of
the effective order parameters appropriate for the continuous DC-OC
and OC-OD transitions.  This hypothesis is quite reasonable, as the
two transition lines are associated with different degrees of freedom:
transitions along the DC-OC line are only driven by the condensation
of bilinear scalar fields, while gauge fields drive the transitions
along the OC-OD line.  The nature of the bicritical point can be
investigated within the Landau-Ginzburg-Wilson (LGW) framework. One
considers the most general scalar $\Phi^4$ theory that is invariant
under O(3)$\oplus$${\mathbb Z}_2$
transformations~\cite{LF-72,FN-74,NKF-74,CPV-03}, i.e.
\begin{eqnarray}
{\cal H}_{\rm LGW} &=& 
\case{1}{2} \Bigl[ ( \partial_\mu \psi)^2  + (
\partial_\mu \varphi)^2\Bigr] + 
\case{1}{2} \Bigl( r_\phi \psi^2  + r_\varphi \varphi^2 \Bigr)
\nonumber\\
&&+
\case{1}{4!} \Bigl[ u_\phi (\psi^2)^2 + u_\varphi (\varphi^2)^2 + 
                           2 w \psi^2\varphi^2 \Bigr] \,,
\label{bicrHH} 
\end{eqnarray}
where the three-component field $\psi$ and the one-component field
$\varphi$ are the coarse-grained O(3) and ${\mathbb Z}_2$ order
parameters. This LGW field theory has been extensively studied in
Refs.~\cite{LF-72,FN-74,NKF-74,CPV-03}.  In the mean-field
approximation ~\cite{LF-72,FN-74,NKF-74}, the model admits a phase
diagram analogous to that sketched in Fig.~\ref{phdiasketch}, with a
bicritical point, where two continuous transition lines (belonging to
the Heisenberg and Ising universality classes) and a first-order
transition line meet.  The nature of the MCP depends on the stability
of the fixed points of the RG flow of the $\Phi^4$ theory
(\ref{bicrHH}). A continuous transition is possible only if the RG
flow admits a stable fixed point.

The RG flow of the model was studied in Ref.~\cite{CPV-03}. Results
were not conclusive for the particular case considered here, leaving
open the possibility of observing a bicritical continuous transition
controlled by the so-called biconical fixed point.  If such fixed
point is stable, the transitions along the DC-OC and OC-OD lines may
be continuous up to the the MCP at $(\kappa_m,J_m)$. In the opposite
case, the continuous transitions along the two lines turn into
first-order transitions before reaching the MCP. It is important to
note that, independently of the stability of the biconical fixed
point, the LGW $\Phi^4$ theory predicts the DC-OD transition line to
be of first order close to the MCP.  As $\kappa$ is increased along
this line, the first-order transition should become weaker and weaker
(the latent heat decreases) as the O(4) continuous transition at
$\kappa=\infty$ is approached, with substantial crossover phenomena
occurring for large values of $\kappa$. Note that, {\em a priori} we
cannot exclude that the first-order transitions turn into continuous
ones for finite values of $\kappa$.

\section{Numerical results for the doubly-charged model}
\label{numres}

We present a numerical study based on MC simulations of the
doubly-charged lattice AH model with $N=2$ and $N=25$.  We consider
cubic lattices of linear size $L$ with periodic boundary
conditions. As in our previous work \cite{PV-19-cpn,PV-19-cah}, we
perform microcanonical and Metropolis updates of the scalar
fields. For the gauge field $\lambda_{{\bm x},\mu}$ we only perform
Metropolis updates: we propose $\lambda_{{\bm x},\mu} \to e^{i\varphi}
\lambda_{{\bm x},\mu}$, choosing $\varphi$ either uniformly around 0
(more precisely, in $0\le |\varphi| \le a$, where $a$ is chosen to
obtain an average acceptance of 30\%) or among the $q-1$ values
$\exp(2 \pi i n/q)$, where $n$ is an integer chosen uniformly among 1,
$\ldots$, $q-1$).

\subsection{Results for $N=2$}

\subsubsection{Results along lines at fixed $J$}

\begin{figure}[tbp]
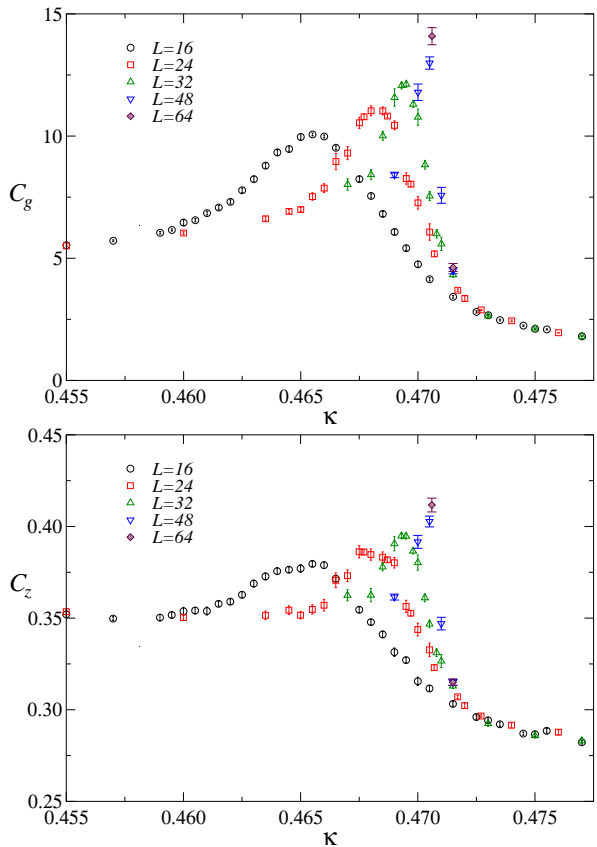

\includegraphics*[scale=\graphicscale,angle=0]{Cg_N2_b0p8.eps}
\includegraphics*[scale=\graphicscale,angle=0]{Cz_N2_b0p8.eps}
\caption{Plot of the specific heats $C_g$ and $C_z$ as a function of 
$\kappa$ for $J = 0.8$ and $N=2$.
}
\label{CP1-b0.8-specificheats}
\end{figure}

\begin{figure}[tbp]
\includegraphics*[scale=\graphicscale,angle=0]{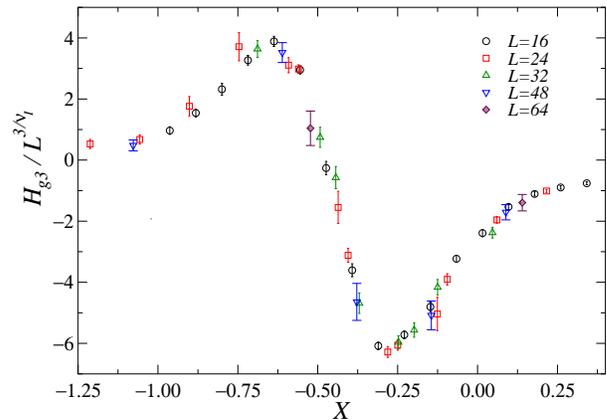}
\caption{Scaling plot of the cumulant $H_{g3}$ as a function of 
$X = (\kappa - \kappa_{c}) L^{1/\nu_I}$, taking 
$\nu_I = 0.629971$ and $\kappa_c = 0.47131$. 
Results for $J = 0.8$ and $N=2$.
}
\label{CP1-b0.8-scalingplotM3}
\end{figure}

\begin{figure}[tbp]
\includegraphics*[scale=\graphicscale,angle=0]{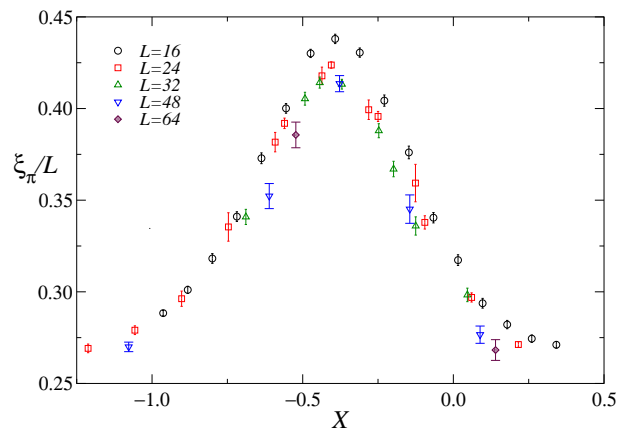}
\caption{Scaling plot of the ratio $\xi_\pi/L$ as a function of $X =
  (\kappa - \kappa_{c}) L^{1/\nu_I}$, taking $\nu_I = 0.629971$ and
  $\kappa_{c} = 0.47131$.  Results for $J = 0.8$ and $N=2$.  }
\label{CP1-b0.8-scalingplotxiPL}
\end{figure}

To begin with, we investigate the behavior of the model along lines
with fixed $J$.  We present a detailed study at $J = 0.8$ (we consider
sizes $16\le L \le 64$) and some additional results for $J = 0.6$,
which, as we shall discuss, is quite close to the MCP.

Since the value $J=0.8$ is larger than $J_c(\kappa = 0) = 0.7102(1)$,
we expect, by increasing $\kappa$, to intersect the OC-OD line, see
Fig.~\ref{phdiasketch}. No other transition is expected if the MCP
satisfies $J_m < 0.8$ (we will verify below that $J_m \lesssim 0.56$).
In Fig.~\ref{CP1-b0.8-specificheats} we show the specific heats $C_g$
and $C_z$ as a function of $\kappa$ at fixed $J = 0.8$. We observe a
peak for $\kappa \approx 0.47$, which confirms the presence of a phase
transition. Quantities related to the observable $Q$ do not show any
singular behavior (although they are expected to be nonanalytic at the
transition), confirming that the transition belongs to the OC-OD
line. The Binder parameter $U$ and the correlation length $\xi$ do not
vary significantly in the transition region. However, they
significantly vary with the size of the system: the values $U \approx
1.0011,1.0003,1.00006$ and $\xi \approx 28.7$, 81, 230 for $L=16,32$,
and 64. Note that $\xi$ increases faster than $L$ with the size of the
box, as expected for an ordered phase---we expect that $\xi\sim L^2$
asymptotically. To characterize the critical behavior, as suggested in
Ref.~\cite{SSNHS-03}, we consider the scaling behavior of the third
moment $H_{g3}$, which is expected to scale as
\begin{equation} 
H_{g3} \approx L^{3/\nu} \left[
   f_3(X) + L^{-\omega} g_3(X) + L^{3-3/\nu} g_{3a}(\kappa)\right],
\label{Hg3-scaling}
\end{equation}
where $X = (\kappa - \kappa_{c}) L^{1/\nu}$, $\omega$ is the 
leading correction-to-scaling exponent and the last term is due to the 
analytic background. For an Ising transition \cite{Hasenbusch-10,KPSV-16}
\begin{equation}
\nu_I = 0.629971(4), \qquad \omega_I = 0.832(6),
\label{expIsing}
\end{equation}
and $3-3/\nu_I = -1.76212(3)$. Corrections decay quite rapidly and the 
analytic background is negligible. Note that this is not the 
case of the specific heats that scale as 
\begin{equation}
C_{g,z} \approx L^{2/\nu-3} \left[ f_C(X) + L^{-\omega} g_C(X) +
  L^{3-2/\nu} g_{Ca}(\kappa)\right]\,.
\end{equation}
In this case, since $3-2/\nu_I \approx -0.17475(2)$, the analytic term
gives rise to quite slowly decaying scaling corrections.

To verify the Ising nature of the transition, we perform nonlinear
fits to Eq.~(\ref{Hg3-scaling}), approximating $f_3(X)$ with a
polynomial of degree $n$ and neglecting all scaling corrections (we
set $g_3(X) = g_{3a}(\kappa) = 0$).  If we take $n=14$ we obtain $\nu
= 0.630(4)$ if we use all data and $\nu = 0.623(10)$ if we only
include data with $L\ge 24$. The results are clearly compatible with
the Ising value $\nu_I$ reported in Eq.~(\ref{expIsing}).  To obtain a
more precise estimate of $\kappa_{c}$ we have repeated the fits fixing
$\nu$ to the Ising value reported above. A fit to all data gives
\begin{equation}
   \kappa_c = 0.47131(2)\,,
\label{estN2k0p8}
\end{equation}
which is in full agreement with the result obtained using only results
with $L\ge 24$, $\kappa_{c} = 0.47128(3)$. In
Fig.~\ref{CP1-b0.8-scalingplotM3} we report the scaling plot of
$H_{g3} L^{-3/\nu}$ versus $X$, using the Ising value $\nu_I$ and the
above-reported estimate of $\kappa_c$.  We observe an excellent
agreement.

To further confirm the presence of long-range correlations, we
consider the correlation length $\xi_\pi$, obtained from the
correlation function $G_\pi$, Eq. ~(\ref{plaope}).  In
Fig.~\ref{CP1-b0.8-scalingplotxiPL} we report $\xi_\pi/L$ as a
function of $X = (\kappa - \kappa_{c}) L^{1/\nu}$, using $\nu_I$ and
the estimate of $\kappa_{c}$ reported in Eq.~(\ref{estN2k0p8}). We
observe a reasonable good scaling, although not perfect, plausibly
because of the corrections due to the analytic background. We also
considered the correlation length $\xi_F$, which, however, is very
small and consistent with zero within errors.

\begin{figure}[tbp]
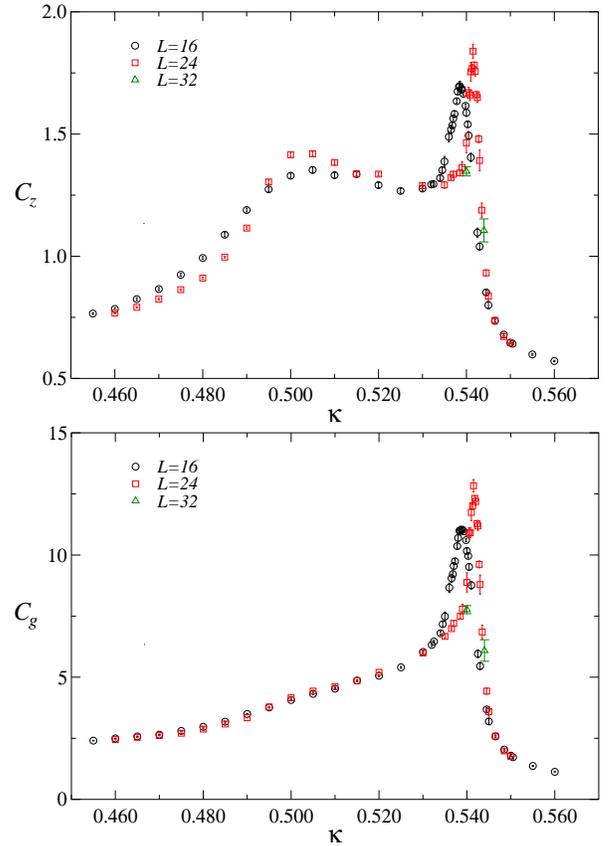

\includegraphics*[scale=\graphicscale,angle=0]{Cz_N2_b0p6.eps}
\includegraphics*[scale=\graphicscale,angle=0]{Cg_N2_b0p6.eps}
\caption{Plot of the specific heats $C_z$ (top) and $C_g$ (bottom) 
as a function of $\kappa$ for $J = 0.6$ and $N=2$.
}
\label{CP1-b0.6-specificheats}
\end{figure}

\begin{figure}[tbp]
\includegraphics*[scale=\graphicscale,angle=0]{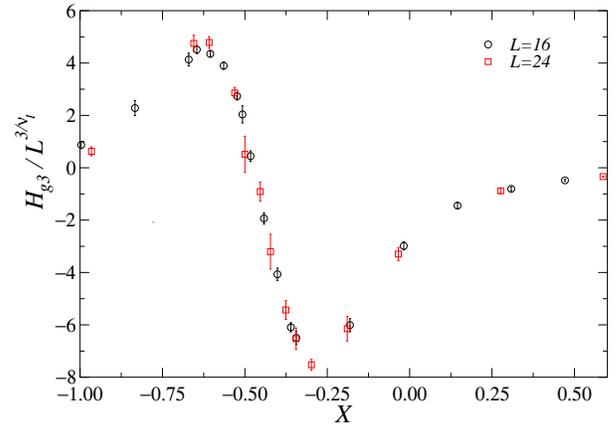}
\caption{Scaling plot of the cumulant $H_{g3}$ as a function of $X =
  (\kappa - \kappa_{c}) L^{1/\nu_I}$, at the Ising transition taking
  $\nu_I = 0.629971$ and $\kappa_c = 0.54472$.  Results for $J = 0.6$
  and $N=2$.  We only include data with $\kappa > 0.53$.  }
\label{CP1-b0.6-scalingplotM3}
\end{figure}

\begin{figure}[tbp]
\includegraphics*[scale=\graphicscale,angle=0]{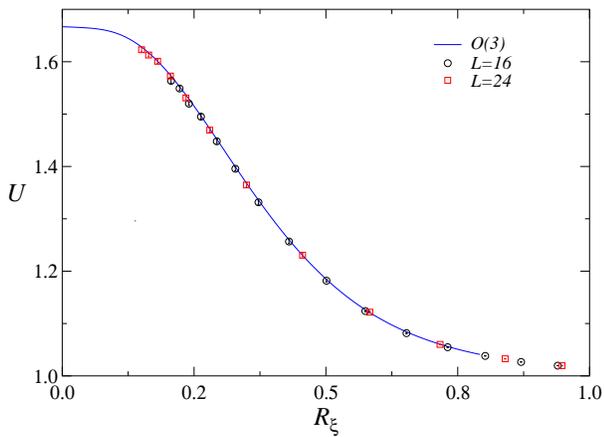}
\caption{Scaling plot of $U$ as a function of $R_\xi$, at the
  Heisenberg transition.  Results for $J = 0.6$ and $N=2$.  We only
  include data with $\kappa < 0.52$.  The continuous line is the
  universal curve $U=F(R_\xi)$ in the standard Heisenberg model
  \cite{footnote-Heisenberg}.  }
\label{CP1-b0.6-scalingplotURxi}
\end{figure}

We have repeated the analysis for $J = 0.6$, which is smaller than
$J_c(\kappa=0) = 0.7102(1)$ \cite{PV-19-cpn} and larger than
$J_c(\kappa=\infty) = 0.233965(2)$~\cite{BFMM-96}. By looking at
Fig.~\ref{phdiasketch}, we see that, by increasing $J$ there are two
possibilities: the line can intersect either one or two transition
lines. The nature of the transition lines depends on the position of
the MCP.  If $J_m < J_c(\kappa=0)$ as in Fig.~\ref{phdiasketch}, there
are two possibilities: if $J > J_m$, by increasing $\kappa$, one would
first cross the DC-OC line and then the OC-OD line; in the opposite
case, $J < J_m$, one would observe only one transition on the DC-OD
line.  On the other hand, if $J_m > J_c(\kappa=0)$, we should observe
a single transition on the DC-OD line. As we discuss below, we find
two transitions, a small-$\kappa$ Heisenberg one---therefore on the
DC-OC line---and a large-$\kappa$ Ising one---therefore on the OC-OD
line.  We can thus conclude that
\begin{equation}
J_m < 0.6 < J_c(\kappa=0)\,, 
\end{equation}
confirming the qualitative correctness of
Fig.~\ref{phdiasketch}.

In Fig.~\ref{CP1-b0.6-specificheats} we report the specific heats as a
function of $\kappa$. The plot of $C_z$ indicates the presence of two
different transitions, one at $\kappa_{c} \approx 0.50$ and a second
one at $\kappa_{c} \approx 0.54$. Note that $C_g$ is little sensitive
to the first transition, which is clearly related to the condensation
of the $Q$ field. The gauge degrees of freedom are only relevant fot
the largest-$\kappa$ transition.

To characterize the large-$\kappa$ transition, we consider the data
corresponding to $\kappa > 0.53$ and fit $H_{g3}$ to $L^{3/\nu}
f_3(X)$ (we take a 14th-order polynomial approximation).  We find $\nu
= 0.625(7)$, consistent with the Ising value (\ref{expIsing}).  If we
fix $\nu$ to the Ising value, we obtain $\kappa_{c} = 0.54472(5)$. The
error here only represents the statistical uncertainty of the fit. The
systematic error due to the scaling corrections is probably larger. In
Fig.~\ref{CP1-b0.6-scalingplotM3} we report the corresponding scaling
plot. The scaling is excellent.  Comparing the estimates of $\kappa_c$
for $J=\infty$ ($\kappa_c \approx 0.381$), $J=0.8$ and $J=0.6$, we see
that $\kappa_c$ decreases with increasing $J$, as sketched in
Fig.~\ref{phdiasketch}.  Moreover, we can derive a lower bound on the
position of the MCP, $\kappa_m \gtrsim 0.55$.

To clarify the nature of the lower-$\kappa$ transition, we plot $U$
versus $R_\xi$, see Fig.~\ref{CP1-b0.6-scalingplotURxi}, considering
only data with $\kappa < 0.52$. Data fall onto a single curve,
compatible with the universal curve computed for the Heisenberg
universality class \cite{footnote-Heisenberg}. To determine the
critical temperature we have performed fits of $U$ and $R_\xi$ to
\begin{equation}
R = f(X)\,, \qquad X = (\kappa - \kappa_{c})L^{1/\nu}\,,
\label{RvsfX}
\end{equation}
taking $\nu$ equal to the Heisenberg value \cite{Hasenbusch-20} $\nu_H
= 0.71164(10)$. Using a polynomial approximation for $f(X)$, we obtain
$\kappa_{c} = 0.4998(2)$ (fit of $U$) and $\kappa_{c} = 0.4996(1)$
(fit of $R_\xi$). Scaling corrections are apparently of the order of
the statistical errors.  We take $\kappa_{c} = 0.4997(3)$ as our final
estimate.

\subsubsection{Results along lines at fixed $\kappa$}  

\begin{figure}[tbp]
\includegraphics*[scale=\graphicscale,angle=0]{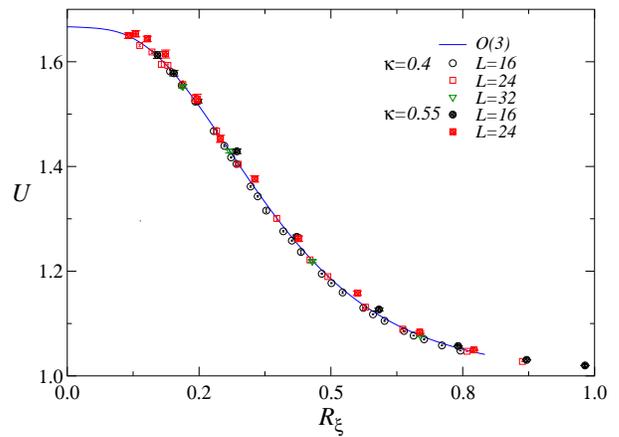}
\caption{Scaling plot of $U$ as a function of $R_\xi$. We report
 data for $N=2$, and 
  $\kappa = 0.4$ and $\kappa=0.55$.  The continuous line is the universal
  curve $U=F(R_\xi)$ in the standard Heisenberg model
  \cite{footnote-Heisenberg}.  }
\label{CP1-bg0.4-scalingplotURxi}
\end{figure}

We now study the behavior of the model by performing simulations at
fixed $\kappa$. We perform simulations at $\kappa = 0.4$, 0.55, and 1.

Since $\kappa_c(J=\infty) < 0.4 < \kappa_m$ (we estimated $\kappa_m
\gtrsim 0.55$), Fig.~\ref{phdiasketch} allows us to predict that, by
increasing $J$ we should first cross the DC-OC line and, subsequently,
the OC-OD line. We focus here on the low-$J$ transition.  The
estimates of $U$ and $R_\xi$ indicate that it occurs for $J\approx
0.65$.  To verify that it belongs to the Heisenberg universality
class, we plot $U$ versus $R_\xi$ and compare the results with the
universal curve computed in the Heisenberg model, see
Fig.~\ref{CP1-bg0.4-scalingplotURxi}. We observe an excellent
agreement.  We have also performed fits of $U$ and $R_\xi$ to
Eq.~(\ref{RvsfX}), with $J$ replacing $\kappa$. We obtain $\nu =
0.721(7)$ and 0.720(3) from $U$ and $R_\xi$, respectively. We observe
tiny deviations from the Heisenberg value \cite{Hasenbusch-20} $\nu_H
= 0.71164(10)$, which are most likely due to scaling corrections.  If
we fix $\nu = \nu_H$, we obtain $J_c = 0.65001(4)$ and $J_c =
0.649708(3)$ from the two fits. Statistical errors are clearly smaller
than the systematic deviations due to the scaling corrections.  A
conservative estimate is $J_c = 0.64985(20)$.

\begin{figure}[tbp]
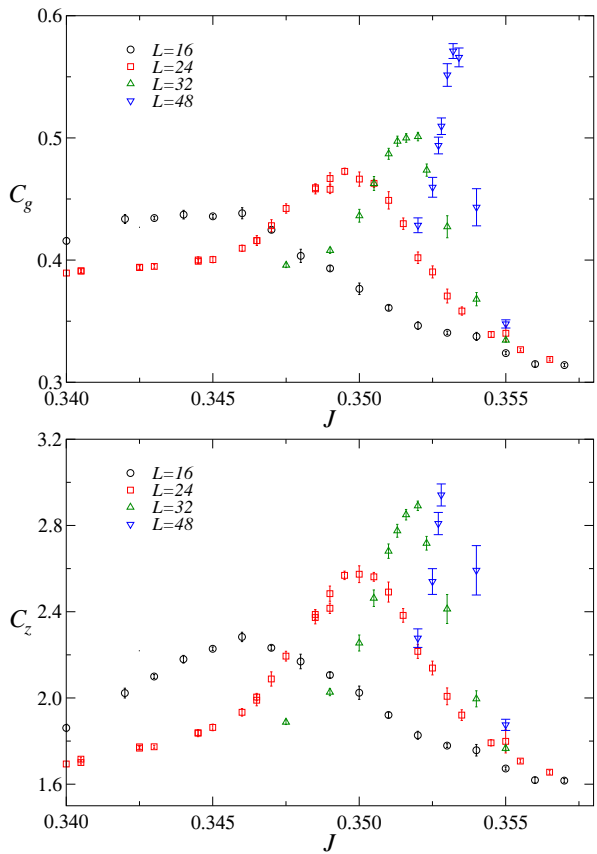

\includegraphics*[scale=\graphicscale,angle=0]{Cg_N2_bg1.eps}
\includegraphics*[scale=\graphicscale,angle=0]{Cz_N2_bg1.eps}
\caption{Plot of $C_g$ (top) and of $C_z$ (bottom) as a function of $J$.
Results for $\kappa = 1$ and $N=2$.
}
\label{CP1-bg1-specificheat}
\end{figure}

\begin{figure}[tbp]
\includegraphics*[scale=\graphicscale,angle=0]{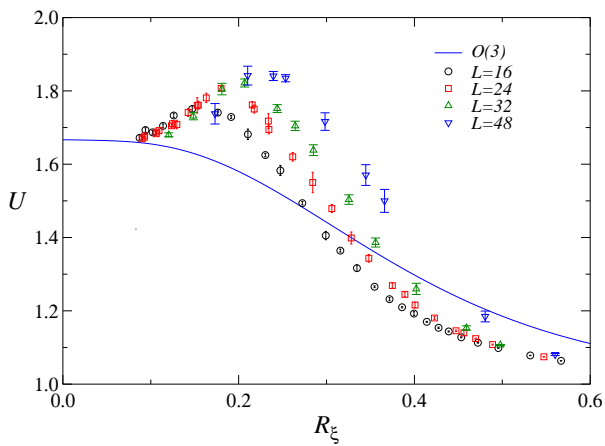}
\caption{Scaling plot of $U$ as a function of $R_\xi$.  Results for
  $\kappa = 1$ and $N=2$.  The continuous line is the universal curve
  $U=F(R_\xi)$ in the standard Heisenberg model
  \cite{footnote-Heisenberg}.  }
\label{CP1-bg1-scalingplotURxi}
\end{figure}

\begin{figure}[tbp]
\includegraphics*[scale=\graphicscale,angle=0]{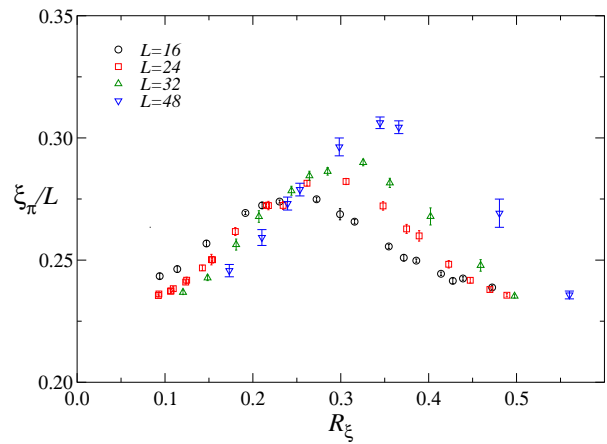}
\caption{Scaling plot of $\xi_{\pi}/L$ as a function of $R_\xi$.
Results for $\kappa = 1$ and $N=2$.
}
\label{CP1-bg1-scalingplotxiPL}
\end{figure}

\begin{figure}[tbp]
\includegraphics*[scale=\graphicscale,angle=0]{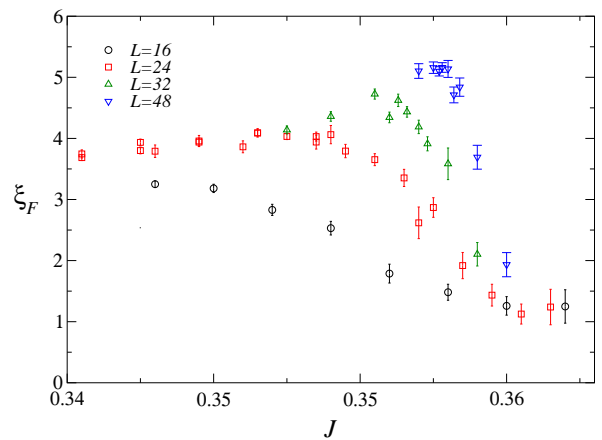}
\caption{Scaling plot of $\xi_F$ as a function of $J$.  Results for
  $\kappa = 1$ and $N=2$.  }
\label{CP1-bg1-xiF}
\end{figure}

We have performed a second set of runs for $\kappa = 0.55$. From the
analysis of the data at fixed $J$ we already know that there is an
Ising transition for $J \approx 0.6$. We wish now to identify the
Heisenberg transition.  We compute $U$ on lattices of size $L=16$ and
24.  We observe a rapid crossover from 5/3 to 1 as $J$ varies from
0.54 and 0.58, allowing us to identify the transition region.  The
results are plotted versus $R_\xi$ in
Fig.~\ref{CP1-bg0.4-scalingplotURxi}. They are very close to the
Heisenberg line, but not exactly on it on the scale of the figure: all
of them are slightly above the scaling curve. We confirm therefore
that the transition is a Heisenberg one, but observe that scaling
corrections are now sizable, as expected since we are now close to the
MCP. As before, we have performed fits of $U$ and $R_\xi$ to
Eq.~(\ref{RvsfX}). We obtain estimates of $\nu$ that are consistent
with the Heisenberg value: $\nu = 0.70(1)$ and 0.72(3) from the
analysis of $R_\xi$ and $U$.  If $\nu$ is fixed to the Heisenberg
value, we estimate $J_c = 0.56419(7)$ and $J_c = 0.56422(12)$,
respectively. Our final estimate is $J_c = 0.5642(2)$.

Finally, we perform runs for $\kappa = 1$, which allows us to study
the behavior along the DC-OD line that connects the MCP with the O(4)
point at $\kappa = \infty$. In Fig.~\ref{CP1-bg1-specificheat} we
report the specific heats $C_z$ and $C_g$, that signal a transition
for $J \approx 0.355$, which is clearly not a Heisenberg one: in the
latter case, since \cite{PV-02} $\alpha < 0$, the specific heat does
not increase with the size $L$.  In Fig.~\ref{CP1-bg1-scalingplotURxi}
we report $U$ versus $R_\xi$.  The data do not scale and, moreover,
$U$ has a maximum whose height apparently increases as $L$
increases. This is the signature of a first-order transition, with a
critical temperature $J_c = 0.354(1)$.  We also report $\xi_\pi/L$
versus $R_\xi$, see Fig.~\ref{CP1-bg1-scalingplotxiPL}. Also the
energy-related correlation length increases with $L$, but with a
finite-size behavior that appears to be unrelated to that of
$\xi$. Finally, in Fig.~\ref{CP1-bg1-xiF} we report $\xi_F$. At all
transition points we have considered above, $\xi_F \approx 0$.  At
this transition $\xi_F$ is sizable, but significantly smaller than
$\xi$ and $\xi_\pi$. Moreover, it does not increase with $L$,
indicating that the modes encoded by this correlation function are not
those that drive the transition.

\subsection{Results for $N=25$}

\begin{figure}[tbp]
\includegraphics*[scale=\graphicscale,angle=0]{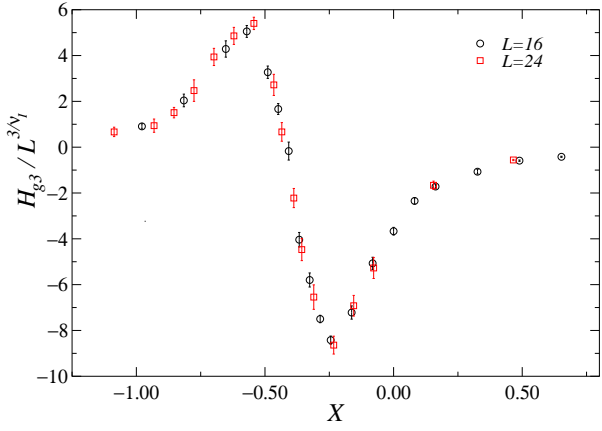}
\caption{ Scaling plot of the cumulant $H_{g3}$ as a function of $X =
  (\kappa - \kappa_{c}) L^{1/\nu_I}$, for $\nu_I = 0.629971$ and
  $\kappa_c = 0.40200$. Results for $J = 0.4$ and $N=25$.  }
\label{CP24-b0.4-scalingplotM3}
\end{figure}

\begin{figure}[tbp]
\includegraphics*[scale=\graphicscale,angle=0]{Cz_N25_bg1.eps}
\includegraphics*[scale=\graphicscale,angle=0]{Cg_N25_bg1.eps}
\caption{Plot of the specific heats $C_z$ (top) and $C_g$ (bottom) 
as a function of $J$ for $\kappa = 1$ and $N=25$.
}
\label{CP24-bg1-specificheats}
\end{figure}

\begin{figure}[tbp]
\includegraphics*[scale=\graphicscale,angle=0]{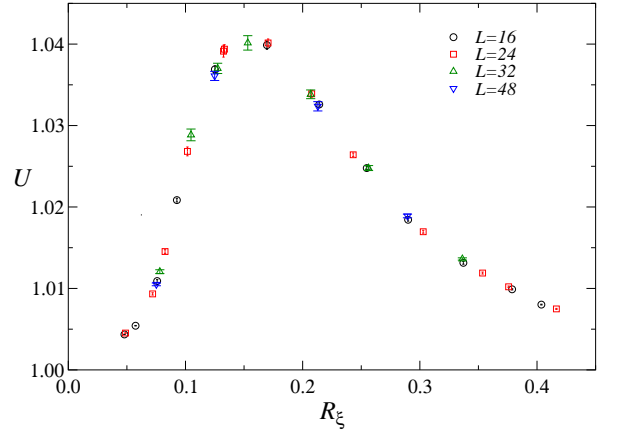}
\caption{Scaling plot of $U$ as a function of $R_\xi$.
Results for $\kappa = 1$ and $N=25$.
}
\label{CP24-bg1-scalingplotURxi}
\end{figure}

\begin{figure}[tbp]
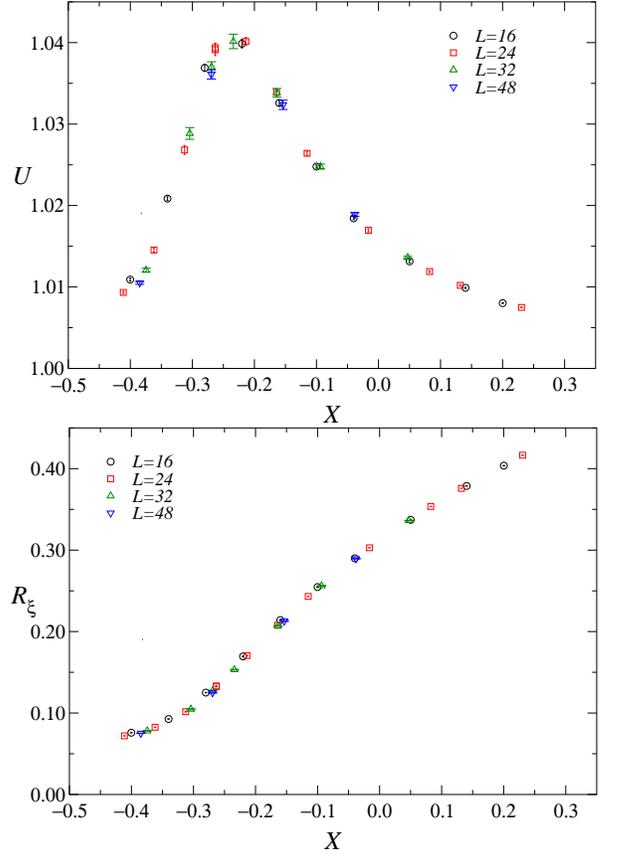

\includegraphics*[scale=\graphicscale,angle=0]{U4_N25_bg1p0.eps}
\includegraphics*[scale=\graphicscale,angle=0]{Rxi_N25_bg1p0.eps}
\caption{ Scaling plot of $U$ (top) and $R_\xi$ versus $X = (J -
  J_{c}) L^{1/\nu}$, for $\nu = 0.815$ and $J_c = 0.29333$. Results
  for $\kappa = 1$ and $N=25$.  }
\label{CP24-bg1-scalingplotU}
\end{figure}

\begin{figure}[tbp]
\includegraphics*[scale=\graphicscale,angle=0]{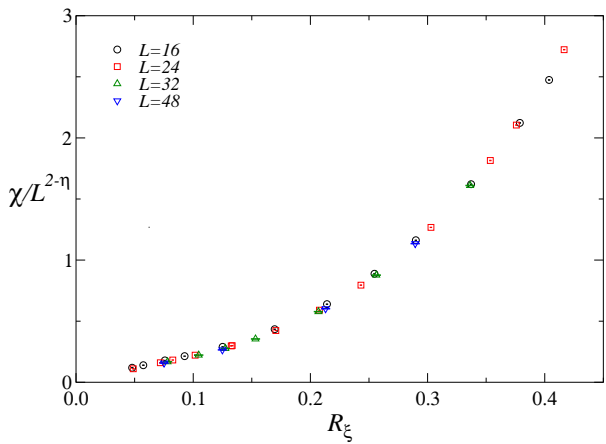}
\caption{ Scaling plot of $L^{-2+\eta_q} \chi $ versus $R_\xi$ for
  $\eta_q = 0.88$. Results for $\kappa = 1$ and $N=25$.  }
\label{CP24-bg1-scalingplotchi}
\end{figure}

We now consider the CP$^{24}$ model and perform an analysis analogous
to the one presented for $N=2$. According to Ref.~\cite{PV-20-largeN},
the compact CP$^{19}$ model has a critical transition at $J_c =
0.353(2)$. Since the transition point in the CP$^{N-1}$ model
decreases with increasing $N$, the CP$^{24}$ model should have a
transition at $J_c \lesssim 0.35$.  Therefore, in order to investigate
the behavior of the model along the OC-OD critical line that connects
the MCP with the Ising transition at $J=\infty$, we have performed a
run at fixed $J = 0.4$. We find that the specific heat $C_g$ has a
clear maximum that increases with $L$ for $\kappa \approx 0.40$, that
we identify as the critical transition.  We analyze the behavior of
the cumulant $H_{g3}$ close the transition, fitting the data to
$L^{3/\nu_I} f(X)$, where $X = (\kappa - \kappa_{c})L^{1/\nu_I}$,
using $\nu_I = 0.629971(4)$ \cite{KPSV-16}. Data scale nicely, see
Fig.~\ref{CP24-b0.4-scalingplotM3}, confirming the Ising nature of the
transition line. We estimate $\kappa_{c} = 0.40200(5)$, where the
error takes into account only the statistical fluctuations of the
fit---the systematic error due to the scaling correction is probably
larger.

We now focus on the DC-OC transition line that connects the CP$^{24}$
transition point with the MCP. We perform runs at $\kappa = 0.2$ on
lattices $L=16$, observing strong hysteresis effects for $J\approx
0.339$.  The transitions on the DC-OC line are therefore of first
order, as the CP$^{24}$ transition for $\kappa = 0$.
 
Finally, we focus on the DC-OD line connecting the MCP with the O(50)
transition point for $\kappa = \infty$. We perform runs for $\kappa =
1$, which should be well above the MCP. On the basis of our results
for the Ising line---the results at fixed $J=0.4$---we expect the MCP
to have $\kappa_m \approx 0.4$-0.5.  In
Fig.~\ref{CP24-bg1-specificheats} we show the specific heats as a
function of $J$. The quantity $C_z$ has a well-defined maximum for $J
\approx 0.29$, which does not increase with the size. This is a clear
sign of a continuous transition with $\alpha < 0$. The specific heat
$C_g$ instead seems to be loosely coupled with the transition: it only
shows short plateaus for the values of $J$ at which $C_z$ has a
maximum. The transition is clearly related to the condensation of the
degrees of freedom associated with the field $\bm z$.  In
Fig~\ref{CP24-bg1-scalingplotURxi} we report $U$ versus $R_\xi$. Data
scale nicely, confirming the continuous nature of the transition.

We now proceed to the computation of the critical indices. To estimate
the correlation-length exponent $\nu$, we first perform fits to
Eq.~(\ref{RvsfX}), where we use a polynomial approximation for $f(X)$.
We obtain $\nu = 0.788(2)$ and $\nu = 0.796(4)$ from the analysis of
$R_\xi$ and $U$, respectively. However, the $\chi^2$ (weighted sum of
the fit residuals) for the fit of $R_\xi$ is large:
$\chi^2/\hbox{DOF}\approx 2.5$, where DOF is the number of degrees of
freedom of the fit. This may indicate that there are sizable scaling
corrections. Therefore, we performed a second set of fits to
\begin{equation}
R = f(X) + L^{-\omega} g(X),
\end{equation}
which includes the leading scaling corrections. The $\chi^2$ of the fit is 
little sensitive to $\omega$ in the range $\omega \gtrsim 0.5$. Varying 
$\omega$ in this range we can estimate 
\begin{equation}
\nu = 0.815(15)\,, \qquad J_c = 0.29333(3)\,.
\end{equation}
We finally estimate the exponent $\eta_q$ defined by the size
dependence of the susceptibility $\chi$ at the critical point: $\chi
\sim L^{2-\eta_q}$. We perform fits to \cite{BPV-20-ncah}
\begin{equation}
\chi = L^{2-\eta_q} [f_\chi(R_\xi) + L^{-\omega}g_\chi(R_\xi)]\,,
\end{equation}
Correspondingly, 
we obtain the estimate
\begin{equation}
\eta_q = 0.88(2)\,.
\end{equation}
The fit is also sensitive to $\omega$ and gives $\omega =
1.1(2)$. This estimate is quite reasonable, since we expect $\omega\to
1$ in the limit $N\to\infty$.

\section{Suppressing  the monopoles in the higher-charge CP$^{N-1}$ model} 
\label{monsup}

In the previous Sections we have discussed the phase diagram of the AH
compact model with $q=2$. As it has been discussed at length in the
literature
\cite{MS-90,MV-04,SBSVF-04,NCSOS-11,NCSOS-13,NCSOS-15,WNMXS-17,PV-20-mfcp},
the critical behavior is supposed to depend on the topology of the
gauge fields and in particular on the presence/absence of monopoles.
In particular, it was shown in our previous work \cite{PV-20-mfcp} on
the $q=1$ monopole-free CP$^{N-1}$ model, that monopole suppression
leads to a different critical behavior. No Heisenberg transition is
observed for $N=2$, while for $N=25$ a novel critical behavior
appears, whose intepretation within the field-theory framework is
still obscure \cite{BPV-20-ncah},

We wish now to investigate the role that monopoles play in
higher-charge systems. As in Ref.~\cite{PV-20-mfcp}, we use the De
Grand-Toussaint \cite{DGT-80} monopole definition and consider the
model in which monopoles are absent. We shall focus on the model with
$\kappa = 0$, that will be referred to as the monopole-free CP$^{N-1}$
(MFCP$^{N-1}$) model. As we shall discuss below, this is will be
enough to conjecture a plausible phase diagram for any $\kappa > 0$.

For $\kappa = 0$, in the absence of the monopole constraint, the model
with charge-$q$ fields is obviously equivalent to the usual
one. However, the introduction of the monopole-suppression constraint
which is applied to the $\lambda$ fields breaks the equivalence and
thus a different behavior is possible. We have performed numerical
simulations, studying two cases, $q=2$ and $q=3$.

\begin{figure}[tbp]
\includegraphics*[scale=\graphicscale,angle=0]{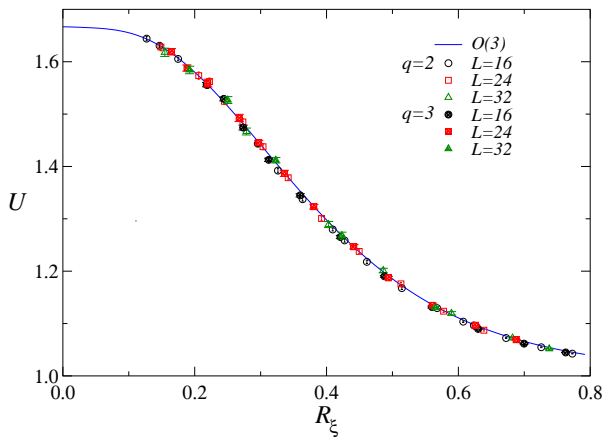}
\caption{Plot of the Binder parameter $U$ versus $R_\xi$ for $q=2$ and
  $q=3$ for the MFCP$^1$ model.  The continuous line is the universal
  curve $U=F(R_\xi)$ in the standard Heisenberg model
  \cite{footnote-Heisenberg}.  }
\label{MFCP1-URxi}
\end{figure}

Let us first consider the MFCP$^1$. We have performed simulations on
systems with $L=16,24,32$.  In Fig.~\ref{MFCP1-URxi} we show our
numerical estimates of $U$ versus $R_\xi$. Data scale nicely, clearly
indicating that scaling corrections are small. Moreover, data fall on
top of the O(3) scaling curve, allowing us to infer that the
transition, for both values of $q$, belongs to the O(3) universality
class, as in the standard CP$^1$ model without monopole
suppression. Apparently the monopole constraint, which is crucial in
determining the critical behavior for $q=1$, plays no role for $q \ge
2$. We also performed fits of the data to
\begin{equation}
   R(\beta,L) = f_R[(J-J_c) L^{1\nu}],
\end{equation}
taking $R = R_\xi$ and $U$, and using a polynomial approximation for
$f_R(x)$. We find that the estimates of $\nu$ are consistent with the
Heisenberg value, $\nu_H \approx 0.711$.  Fixing $\nu$ to the
Heisenberg value \cite{Hasenbusch-20} $\nu_H = 0.71164(10)$, we can
estimate the critical temperature: $J_c = 0.7074(3)$ for $q=2$ and
$J_c = 0.6980(6)$ for $q=3$.  These two estimates are very close to
the transition temperature for the standard CP$^1$ model, $J_c =
0.7102(1)$, while they differ significantly from the $q=1$ MFCP$^1$
value, $J_c = 0.4605(3)$. It is quite clear that the monopole
constraint is not effective for $q\ge 2$.  To verify the irrelevance
of monopole suppression we have also performed simulations with $q=2$
and larger values of $N$. For $N=10,20$ we observe strong first-order
transitions, that are completely analogous to those observed in the
standard CP$^{N-1}$ model. Apparently, monopole suppression is not
effective for any value of $N$.

The results obtained for $\kappa =0$ allow us to conjecture that
monopole suppression does not play any role for any $\kappa > 0$.
Indeed, the introduction of the interaction term $H_g$ defined in
Eq.~(\ref{Hamg}) leads to a reduction of the number of monopoles as
$\kappa$ increases.  Thus, we expect the role of monopole suppression
to be maximal for $\kappa = 0$. Therefore, the irrelevance for $\kappa
= 0$ leads us to predict that they irrelevant for any $\kappa > 0$.

\section{Conclusions} \label{conclu}

We have considered 3D lattice AH models with compact gauge fields and
multicomponent complex scalar fields with integer charge $q$. We have
discussed the dependence on the charge $q$ of the phase diagram and
the nature of the phase transitions.

The resulting phase diagram for $q=2$---but we expect a qualititively
similar diagram for any $q\ge 2$---and unit-length scalar fields
(London limit) is sketched in Fig.~\ref{phdiasketch}.  There are three
phases, separated by three different transition lines meeting at a
MCP. There is one phase (DC phase) in which the scalar fields are
disordered---the SU($N$) global symmetry is unbroken---and
single-charge external particles are confined---Wilson loops obey the
area law for large sizes. In the other two phases (OC and OD) the
SU($N$) symmetry is broken. They differ in the behavior of
single-charge external particles: they are confined for small $\kappa$
(OC phase), deconfined for large $\kappa$ (OD phase).

The nature of the transition lines depends on the number of
components.  For $N=2$ the transitions along the DC-OC line are
continuous, belonging to the O(3) vector universality classes. Along
this line the scalar-field degrees of freedom order, the SU($N$)
symmetry breaks, while gauge-field modes play no role. The transitions
along the OC-OD line are continuous and belong to the Ising
universality class. The transitions are driven by the gauge degrees of
freedom, while the scalar field plays no role.  Finally, first-order
transitions are observed along the DC-OD line, ending to the O(4)
critical point for $\kappa\to\infty$.  The phase diagram substantially
differs from that observed for the compact AH model with single-charge
scalar fields~\cite{PV-19-cah}, whose phase diagram has two phases,
differing only in the behavior of the gauge-invariant scalar-field
correlations; they are separated by a line of continuous transitions
belonging to the O(3) vector universality class~\cite{PV-19-cah},
where the gauge correlations do not play any role.

To characterize the large-$N$ behavior of the model, we have also
reported a numerical study for $N=25$. The transitions along the OC-OD
line are continuous and belong to the Ising universality class, like
the case $N=2$.  The behavior along the DC-OC and OC-OD line instead
differs.  Indeed, the DC-OC transitions are of first order, while the
DC-OD transitions are continuous. We have determined the
correlation-length exponent $\nu$ and the exponent $\eta_q$ that
characterizes the critical behavior of the order parameter $Q_{\bm x}$
defined in Eq.~(\ref{qdef}), on the DC-OD line.  We find
$\nu=0.815(15)$ and $\eta_q=0.88(2)$.  Interestingly, they are in
quantitative agreement with the exponents obtained along the
Coulomb-Higgs transition line in the AH model with noncompact gauge
fields~\cite{BPV-20-ncah}: $\nu=0.802(8)$ and $\eta_q=0.883(7)$.  In
Ref.~\cite{BPV-20-ncah} we conjectured that the Coulomb-Higgs
transitions in the noncompact AH model belong to the universality
class associated with the stable fixed point of the multicomponent AH
field theory \cite{HLM-74,MZ-03,IZMHS-19}, cf. Eq.~(\ref{abhim}).
Indeed, the numerical results were in excellent agreement with the
$1/N$ predictions computed in the continuum AH
model~\cite{HLM-74,IKK-96}: $\nu \approx 0.805$ and $\eta_q \approx
0.870$.  This comparison leads us to conjecture that also the critical
behavior of the large-$N$ continuous DC-OC transitions belongs to the
universality class associated with the stable 3D fixed point of the AH
field theory.

The analogies between the phase diagram of compact AH models with
doubly-charged multicomponent scalar fields and that of the noncompact
AH models~\cite{BPV-20-ncah}, and, in particular, the correspondence
of the DC-OD critical behavior with the AH field theory, may appear
quite unexpected, and certainly deserves further investigation. We
finally note that the noncompact formulation of the AH theory is
recovered in the $q\to\infty$ limit of the compact formulation, with
an appropriate rescaling of the gauge couplings, $\kappa_{\rm nco} =
\kappa/q^2$ ($\kappa_{\rm nco}$ in the inverse gauge coupling in the
noncompact model). Thus, it is tempting to conjecture that the phase
diagram and, in particular, the nature of the transitions along the
DC-OD line, is the same for any $q \ge 2$.


\begin{thebibliography}{99}
\bibitem{footnote-Heisenberg} 
We report here an accurate expression 
(deviations are assumed to be well below 0.5\%) of the universal 
curve $U = F(R_\xi)$ for the Heisenberg universality class: 
$F(x) = 5/3 + x\ (3.0263535 + 23.139470 x) (1-e^{-15x}) 
- 47.838890 x^2  + 58.489668 x^3  - 67.020681 x^4  + 38.408855 x^5 
- 8.8557348 x^6$.
\end{thebibliography}
\end{document}